\documentclass[pdflatex,sn-mathphys-num]{sn-jnl}
\usepackage{graphicx}%
\usepackage{multirow}%
\usepackage{amsmath,amssymb,amsfonts}%
\usepackage{amsthm}%
\usepackage{mathrsfs}%
\usepackage[title]{appendix}%
\usepackage{xcolor}%
\usepackage{textcomp}%
\usepackage{manyfoot}%
\usepackage{booktabs}%
\usepackage{algorithm}%
\usepackage{algorithmicx}%
\usepackage{algpseudocode}%
\usepackage{listings}%
\usepackage{float}

\theoremstyle{thmstyleone}%
\theoremstyle{thmstyletwo}%

\theoremstyle{thmstylethree}%

\begin{document}

\title[Article Title]{A low-volume cold-atom microwave clock with an additively manufactured cavity}

\author[1]{\fnm{Samuel A.} \sur{Smith}} \equalcont{These authors contributed equally to this work.}
\author[1]{\fnm{Sean} \sur{Dyer}}\equalcont{These authors contributed equally to this work.}
\author[1]{\fnm{Alan} \sur{Bregazzi}}
\author[1]{\fnm{Rajnandan} \sur{Choudhury Das}}
\author[1]{\fnm{Oliver S.} \sur{Burrow}}
\author[2]{\fnm{Christoph} \sur{Affolderbach}}
\author[2]{\fnm{Gaetano} \sur{Mileti}}
\author[1]{\fnm{Erling} \sur{Riis}}
\author*[1]{\fnm{Paul F.} \sur{Griffin}}\email{paul.griffin@strath.ac.uk}

\affil[1]{\orgdiv{Department of Physics, SUPA}, \orgname{University of Strathclyde}, \orgaddress{\street{107 Rottenrow E}, \city{Glasgow}, \postcode{G4 0NG}, \country{United Kingdom}}}

\affil[2]{\orgdiv{Institut de Physique, Laboratoire Temps-Fréquence}, \orgname{Université de Neuchâtel}, \orgaddress{\street{Avenue de Bellevaux 51}, \city{Neuchâtel}, \postcode{2000}, \country{Switzerland}}}

\abstract{We present the performance of a compact microwave interrogated atomic clock based on laser cooled $^{87}$$\rm{Rb}$ atoms. The clock utilizes a micro-fabricated grating magneto-optical trap (gMOT) to cool and trap atoms, and an additively manufactured microwave cavity for microwave interrogation. We demonstrate a signal-to-noise ratio (SNR) of $370$ at a 1~Hz bandwidth and mid-term instability of \mbox{$6.7\times10^{-14}$}~at an averaging time of ~\mbox{$\tau=3 \times 10^{4}$}~s. In addition, a low cost mu-metal foil has been implemented on the vacuum chamber to reduce the quadratic Zeeman shift which was previously limiting stability beyond \mbox{$\tau = 10$}~s.}
\maketitle

\section{Introduction}\label{sec1}
Over the past several decades, there has been increasing effort to reduce the size, weight, and power (SWaP) requirements of clock systems while maintaining high timing performance. The drive for compact and deployable atomic clocks is critical for applications requiring precise timing in many fields such as telecommunications, electrical grid synchronization, and navigation, particularly in GNSS-denied environments~\cite{blackettreview,cuntz2012,flytkjaer2023}. Microwave atomic clocks based on pulsed optical pumping (POP) have demonstrated excellent short- and medium-term frequency stability in thermal vapour systems~\cite{batori2022,gozzelino2023}. In the POP scheme, the optical preparation and detection stages are temporally separated from the microwave interrogation, reducing light-induced frequency shifts while enabling high-contrast Ramsey interrogation. Although POP clocks based on thermal vapour achieve excellent short- and medium-term stability, cold-atom clocks offer complementary advantages, including reduced atomic motion, which enables longer interrogation times, and the absence of buffer-gas-induced shifts, making them attractive for improved long-term holdover and frequency accuracy~\cite{moreno2018,carle2023,walby2026}. Combining POP interrogation with a compact cold-atom platform offers a promising route towards a microwave clock with both high short-term stability and robust long-term performance. Several groups have demonstrated deployable cold-atom microwave clocks~\cite{esnault2011,esnault2008,deng2024,lee2021} along with the first commercially available systems now on the market \cite{ascarrunz2018,8597468}. 

The cold-atom microwave clock presented here is being developed with the goal of achieving a highly compact architecture that is compatible with scalable manufacturing and deployable operation while maintaining performance comparable to existing state-of-the-art systems. The clock is based on laser-cooled $^{87}\mathrm{Rb}$ atoms and employs a double-resonance architecture in which optical transitions are used to perform trapping, cooling, state preparation, and detection, while the ground state is interrogated by microwaves using Ramsey spectroscopy during free-fall~\cite{ramsey1950,sesko1989,kasevich1989,Sesko89}. The additively manufactured microwave cavity has a volume of approximately $67~\mathrm{cm}^3$~\cite{bregazzi2024lowvolume,batori2023} and is housed within a vacuum chamber whose volume could be substantially reduced in future generations of the clock. The compact cavity, combined with a micro-fabricated grating magneto-optical trap (gMOT)~\cite{nshii2013,mcgilligan2015,mcgilligan2016,mcgilligan2017}, enables a more compact and simplified clock architecture by reducing the number of optical components, easing alignment requirements, and simplifying microwave cavity manufacturing.

\section{Experimental Setup}\label{sec2}
The experimental sequence used during clock operation is summarised below and illustrated in Fig.~\ref{cavityAndEnergyLVLS}.

\subsection{Trapping and Cooling}\label{subsec2}
A collimated cooling and trapping beam enters the vacuum chamber through a viewport located above the microwave cavity. This beam has a $1/e^2$ radius of 20~mm and is incident on a grating chip positioned at the base of the microwave cavity, as shown in Fig.~\ref{cavityAndEnergyLVLS} a). This geometry reduces the number of optical components and optical alignment degrees of freedom compared with a conventional six-beam MOT, while maintaining efficient laser cooling. Cooling and trapping light is applied for 500~ms to load the MOT from background vapour coming from resistively heated rubidium dispensers, resulting in a typical atom number of $3 \times 10^6$ atoms, currently limited by the available rubidium vapour pressure. Repump light is generated using an electro-optic modulator (EOM), with the repumping sideband containing approximately 10\% of the cooling carrier power.

The magnetic quadrupole field for trapping is generated by in-vacuum anti-Helmholtz coils wound directly around an aluminium mount which houses the microwave cavity. Following the MOT load, an 8~ms polarization-gradient cooling (PGC) phase is implemented, cooling atoms to sub-Doppler temperatures \cite{mcgilligan2016}.

During this stage, the peak intensity of the cooling beam is ramped from $\approx 5 I_{\mathrm{sat}}$ to $\approx 0.5 I_{\mathrm{sat}}$, where $I_{\mathrm{sat}} = 1.67~\mathrm{mW/cm^2}$ is the saturation intensity, while the laser frequency is further red-detuned from $2\Gamma$ to $5\Gamma$, where $\Gamma = 2 \pi\times6.07~\mathrm{MHz}$ is the natural linewidth of the $^{87}$Rb D$_2$ transition.
 At the conclusion of sub-Doppler cooling, the temperature of the ensemble is measured to be below $20\, \mu \mathrm{K}$, with approximately 20$\%$ of the atoms lost during this stage. This atomic temperature is determined using a time-of-flight (TOF) imaging system through a 4~mm aperture on the side of the microwave cavity.

\subsection{State Preparation}\label{subsec3}
Following the sub-Doppler cooling stage, atoms are prepared in the desired clock state via optical pumping. The optical pumping beam enters the microwave cavity through the 4~mm aperture on the side of the cavity, orthogonal to the cooling beam axis. The beam is retro-reflected by a mirror positioned on the opposite side of the vacuum chamber. The linear polarized beam has a 7~mm $1/e^2$ diameter, providing a nearly uniform intensity profile across the atomic ensemble. Optical pumping is performed on the $F=2 \rightarrow F'=2$ transition with an equal power sideband power driving the  $F=1 \rightarrow F'=2$ repump transition. This 2~ms on-resonance optical pumping pulse prepares more than 95\% of the atomic ensemble in the $\left|F=2,\ m_{\mathrm{F}}=0\right\rangle$ clock state~\cite{duan2017}, thereby increasing signal contrast.

\subsection{Microwave Interrogation}\label{subsec4}
After state preparation, the atoms are interrogated on the clock transition
\(\left|F=2,\,m_\mathrm{F}=0\right\rangle \rightarrow \left|F=1,\,m_\mathrm{F}=0\right\rangle\)
using the TE\(_{011}\)-like mode of the additively manufactured loop-gap resonator (LGR) cavity, which provides an excellent field orientation factor of $\xi = 97\%$~\cite{batori2023}. The cavity has a measured quality factor of \(Q_\mathrm{c}=360\) and a 19~MHz linewidth near the \({}^{87}\mathrm{Rb}\) clock frequency~\cite{batori2023}. The microwave field is generated by a Keysight E8257D analogue signal generator and coupled to the cavity by a coupling-loop on the base of the cavity. A microwave power of $\mathrm{-12}$~dBm at the cavity input and a pulse duration of 200~\(\mu\mathrm{s}\) realise a \(\pi/2\) pulse. Clock operation is performed using Ramsey interrogation, with two \(\pi/2\) pulses separated by a free-evolution time \(T_\mathrm{R}\). Scanning the microwave frequency across resonance produces Ramsey fringes, which provide the frequency discriminator used to stabilize the clock.

\subsection{Detection}\label{subsec5}
\begin{figure}[H]
    \centering
    \includegraphics[width=1\linewidth]{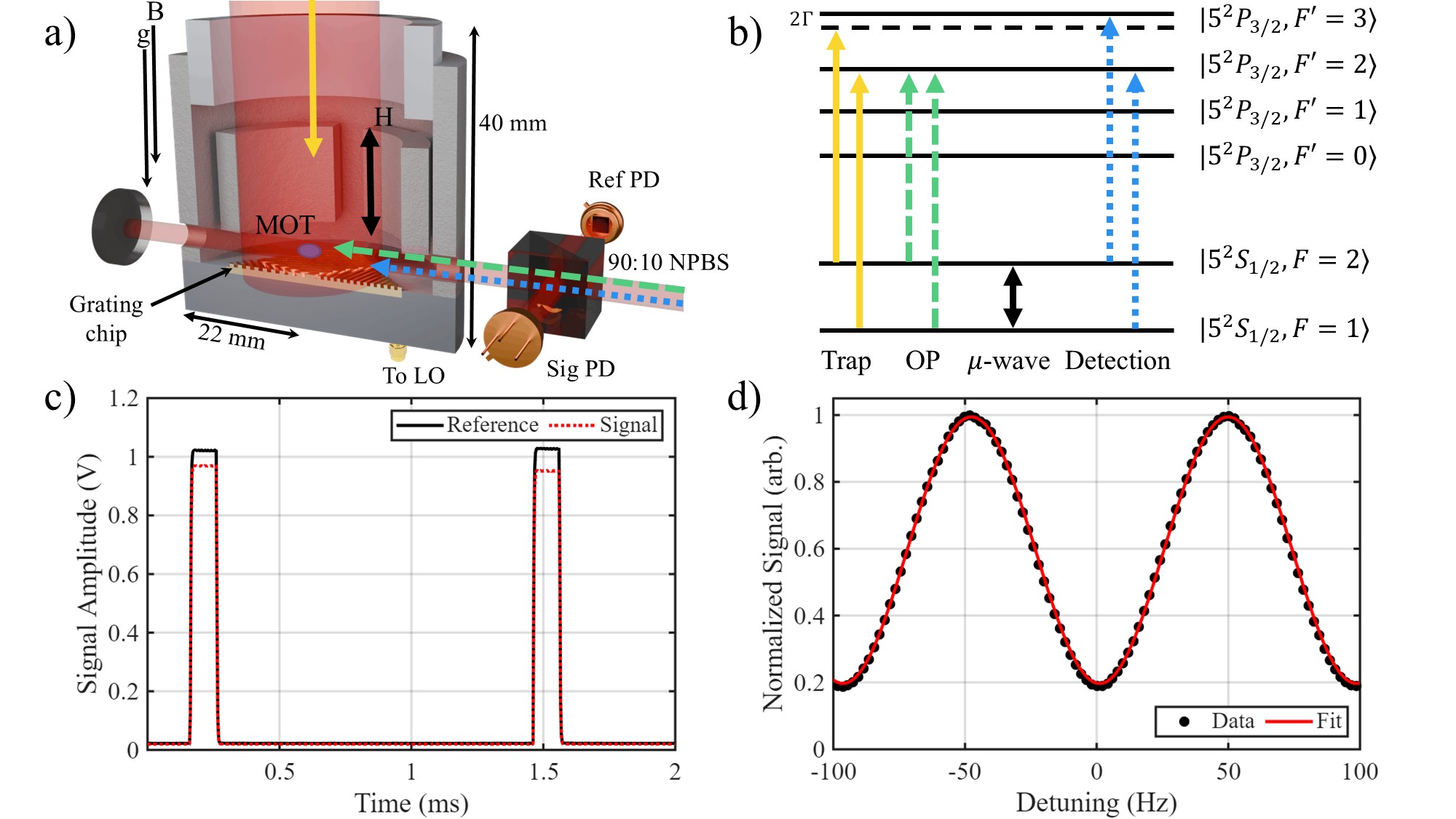}
\caption{a) Simplified diagram of the additively manufactured microwave cavity. The single trapping and cooling beam (solid yellow arrow) is incident on the gMOT chip, forming the cold-atom cloud. The MOT is formed in the region where the atoms can interact with the optical pumping (dashed green arrow) and detection (dotted blue arrow) light. The B field (black arrow) beside the cavity indicates the direction of the bias field applied during clock operation, which is collinear with the H field of the cavity. The black arrow labelled g indicates the direction of gravity. b) Energy-level diagram showing the transitions used in the double-resonance experimental sequence. Solid yellow arrows: trapping light; dashed green arrows: optical pumping light; solid black double arrow: microwave interrogation; dotted blue arrows: detection light. c) Raw detection pulses. Red dotted line: signal photodiode; solid black line: reference photodiode. The repump pulse is not visible because the repump beam propagates orthogonally to the detection axis. d) Typical normalized central Ramsey fringe (black dots) with fit (red line).}
    \label{cavityAndEnergyLVLS}
\end{figure}

Following the microwave interrogation, state readout is performed using absorption detection with a linearly polarized probe beam resonant with the \(\left|F=2\right\rangle \rightarrow \left|F'=3\right\rangle\) closed cycling transition. Absorption detection is employed because the compact microwave cavity geometry limits the optical access required for fluorescence imaging. The probe beam is split by a 90:10 beam splitter before entering the vacuum chamber. The 10\% arm probes the atoms and is detected by a signal photodiode, while the remaining 90\% is directed to a reference photodiode for common-mode intensity-noise rejection as shown in Fig.~\ref{cavityAndEnergyLVLS}.

First, the resonant probe light is pulsed on for 100~$\mu\mathrm{s}$ to measure the population in the \(F=2\) state. The repump beam is then switched on for 1~ms to transfer the remaining atoms into the \(F=2\) state. Finally, the probe light is pulsed on again for 100~$\mu\mathrm{s}$ to measure the total number of atoms interrogated in each experimental cycle. This three-pulse detection sequence enables normalization of the measured \(F=2\) population to the total atom number, reducing the sensitivity of the clock signal to shot-to-shot atom number fluctuations, measured to be approximately \(2.1\%\) RMS.

\section{Results}\label{sec3}
\subsection{Short-term stability}
The short-term stability of an atomic clock is determined by the signal-to-noise ratio (SNR), the linewidth of the clock transition and the experimental cycle time. It can be estimated as

\begin{equation}
\sigma_y(\tau) =
\frac{1}{\pi}
\frac{\Delta \nu}{\nu_0}
\frac{1}{\mathrm{SNR}}
\sqrt{\frac{T_\mathrm{c}}{\tau}},
\label{eq:short_term_stability}
\end{equation}
where $\Delta \nu$ is the Ramsey fringe linewidth, $\nu_0$ is the clock transition frequency, $T_\mathrm{c}$ is the experimental cycle time and $\tau$ is the averaging time.

\begin{figure}[H]
    \centering
    \includegraphics[width=1\linewidth]{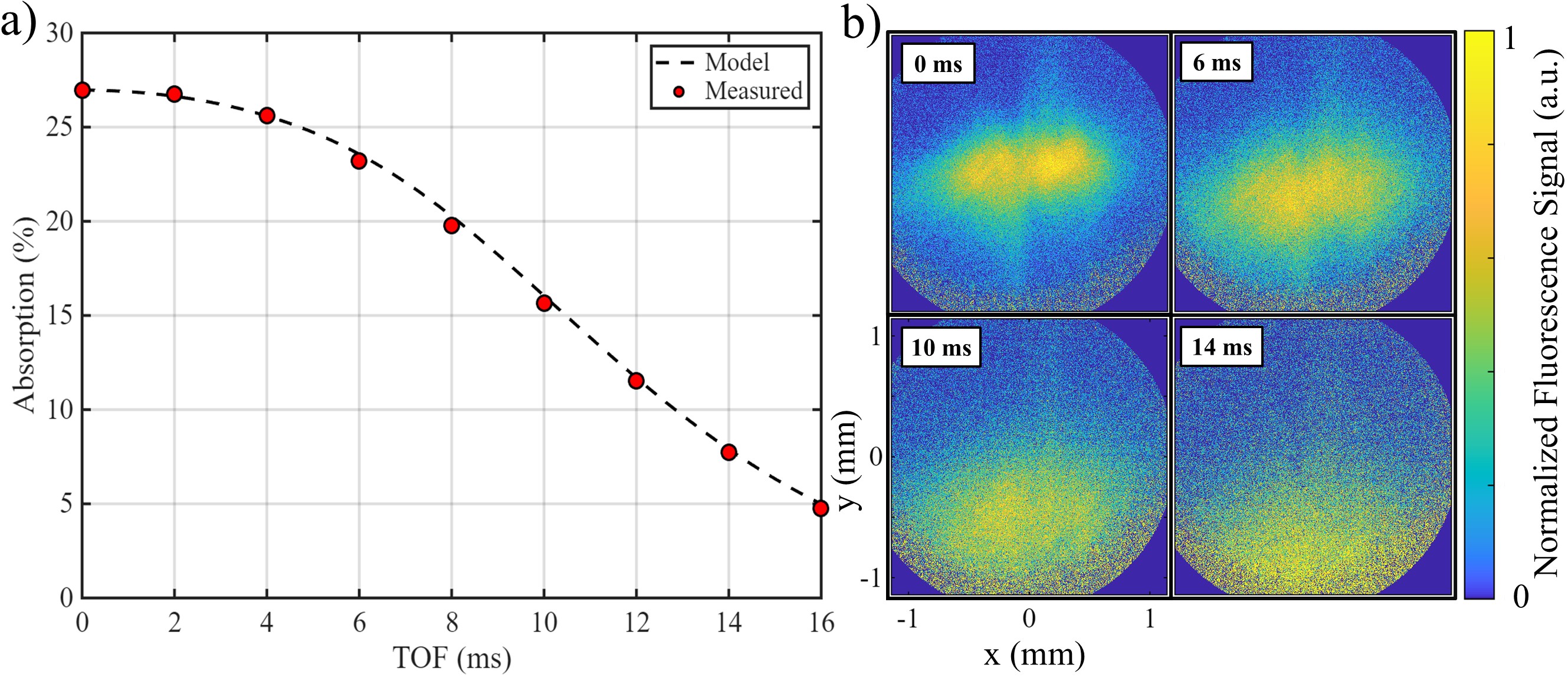}
    \caption{a) Measured absorption of a double-pass resonant probe beam (red dots) as a function of the free-fall time between microwave interrogation and detection. The model (black dashed line) describes atom loss due to free-fall and thermal expansion, accounting for the cloud’s initial position, temperature, and the readout beam intensity profile.
 b) MOT imaging at four different TOF's after state preparation: 0 ms, 6 ms, 10 ms, 14 ms.}
    \label{fig:TOF_ABS_SNR}
\end{figure}

In the present system, the achievable SNR is restricted by two main effects. The first is the reduction in absorption signal during the free-evolution time. The detection beam is aperture-limited by the optical access through the microwave cavity and vacuum system, giving an effective probe diameter of 4~mm, with this dimension chosen to maintain good microwave field homogeneity across the atomic ensemble~\cite{batori2023}. As the time between state preparation and readout is increased, the cold-atom ensemble falls under gravity and thermally expands causing atoms to exit the probe volume. This reduces the absorption signal at readout, indicating a reduced number of atoms within the detection region as shown in Fig.~\ref{fig:TOF_ABS_SNR}.

\begin{figure}[H]
    \centering
    \includegraphics[width=0.7\linewidth]{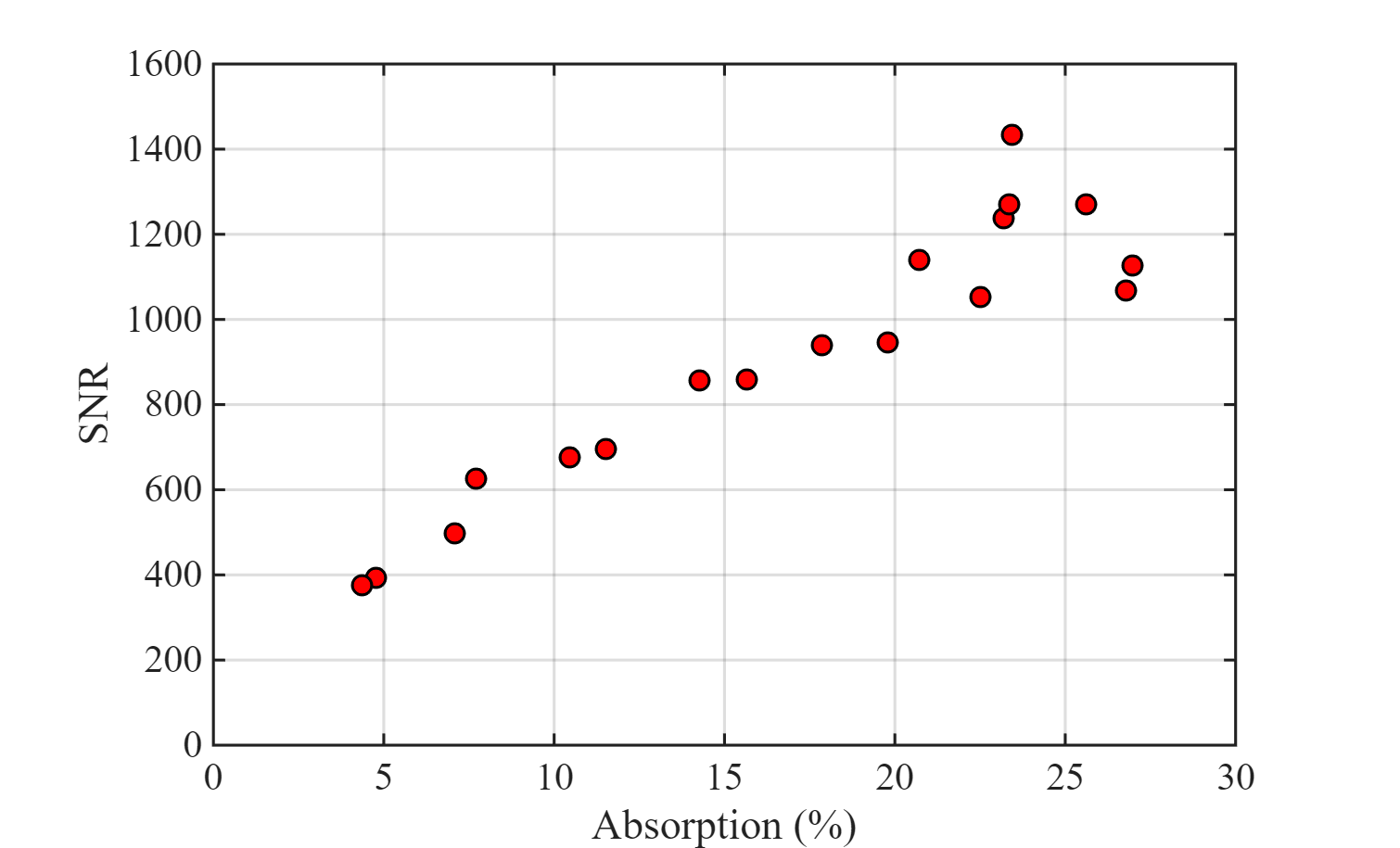}
    \caption{Measured SNR on the side on the Rabi pedestal vs absorption during readout.}
    \label{fig:SNRvsAbsorption}
\end{figure}

The effect of atoms falling out of the detection region on the measured SNR is characterised using Rabi interrogation, implemented with a single microwave $\pi$ pulse. Following state preparation into the $\left|F=2,\ m_{\mathrm{F}}=0\right\rangle$ state, the atoms are interrogated at three microwave frequencies corresponding to the top, bottom, and side of the Rabi resonance. A variable delay of 0-16~ms is introduced between the microwave interrogation and detection, varying the detected absorption without changing the microwave linewidth or experimental cycle time. At each delay, the measurement is repeated over 200 shots to determine the SNR. The signal is defined as the difference in the mean measured amplitudes between the top and bottom of the Rabi resonance, while the noise is given by the standard deviation of the repeated measurements at the side of the resonance. At the shortest delay before detection, the measured absorption is largest. For a cold-atom absorption of 27\%, an SNR of approximately 1100 is measured, as shown in Fig.~\ref{fig:SNRvsAbsorption}.  At longer times of flight, fewer atoms remain within the detection region, reducing the measured absorption and achievable SNR. This measurement demonstrates that the aperture-limited detection region restricts the achievable SNR at longer free-evolution times.

\begin{figure}[H]
    \centering
    \includegraphics[width=0.7\linewidth]{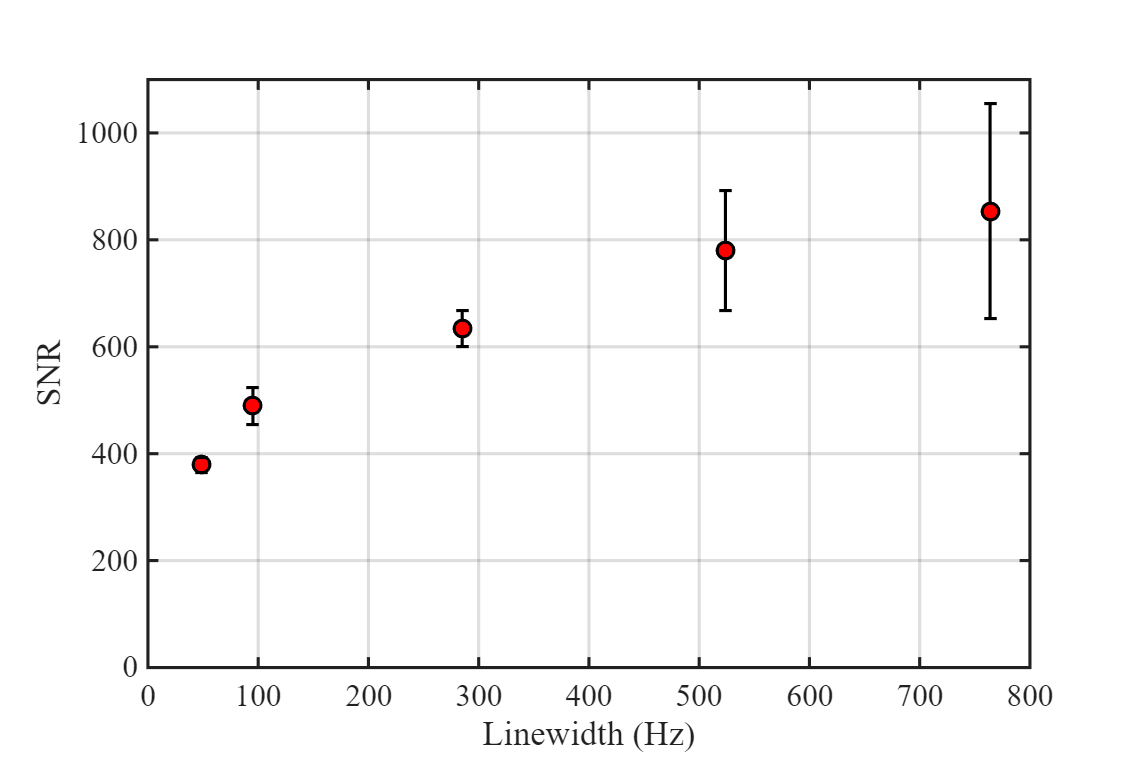}
    \caption{Measured SNR as a function of the central Ramsey fringe linewidth obtained by varying the Ramsey interrogation time while maintaining a constant readout absorption of 15\%.}
    \label{fig:Linewidth}
\end{figure}

The second limitation is observed under Ramsey interrogation. Here, the delay between state preparation and detection is fixed at 10~ms, giving a constant absorption of $15\%$ at readout. The Ramsey interrogation time is varied between 400~\(\mu\mathrm{s}\) and 10~ms, changing the linewidth of the central Ramsey fringe while maintaining a fixed detected absorption. As the Ramsey time is increased, the fringe linewidth narrows and the measured SNR decreases, as shown in Fig.~\ref{fig:Linewidth}. Because the detected absorption is fixed in this measurement, the reduction in SNR cannot be attributed to atom loss from the detection region. Instead, it indicates an additional noise contribution associated with Ramsey interrogation. The source of this is likely local-oscillator noise, which is converted into population noise by the Ramsey discriminator and appears as amplitude noise in the detected signal~\cite{walby2026}. This reduces the effective SNR of the Ramsey discriminator and limits the achievable short-term stability.

Using the measured Ramsey linewidth of 50~Hz, an SNR of 370, and an experimental cycle time of 550~ms, Eq.~\ref{eq:short_term_stability} yields an estimated short-term stability of $4.7 \times 10^{-12}/\sqrt{\tau}$. The high-SNR measurements were obtained at an Rb pressure of approximately $10^{-8}$~mbar. However, this pressure could not be maintained over long timescales due to limitations of the Rb source, preventing a sustained clock measurement under the same operating conditions.

\subsection{Second Order Zeeman Shift and Instability}\label{sec7}
The first generation of this clock was primarily developed to demonstrate the integrated gMOT and additively manufactured microwave cavity architecture, rather than to optimize clock stability. Consequently, magnetic shielding was not incorporated into the original design. As reported previously in~\cite{bregazzi2024lowvolume}, fluctuations in the ambient magnetic field produced corresponding fluctuations in the quadratic Zeeman shift, limiting the medium-term fractional frequency stability to approximately $2\times10^{-12}$.

The magnetic field noise at the atoms is minimized by triggering the experimental sequence synchronously with the AC line and residual fields are minimized using shim coils. However, at low quantisation fields, fluctuations in the transverse magnetic field can tilt the quantisation axis, resulting in reduced Ramsey fringe contrast. Consequently, a relatively large quantisation field of 134~mG was previously required to maintain a fringe contrast of approximately 90\%.

To mitigate this limitation, the vacuum chamber was enclosed using Aaronia MagnoShield Flex magnetic shielding. The thin flexible magnetic-shielding sheets were arranged as a single-layer, cubic enclosure surrounding the external vacuum chamber and shim coils, with openings cut only where required for optical access and vacuum flanges. The shielding provided a measured magnetic-field attenuation factor of approximately four at the position of the atoms. With the additional shielding, the quantisation field could be reduced from 134~mG to approximately 20~mG while maintaining the same Ramsey fringe contrast.

\begin{figure}[H]
    \centering
    \includegraphics[width=1\linewidth]{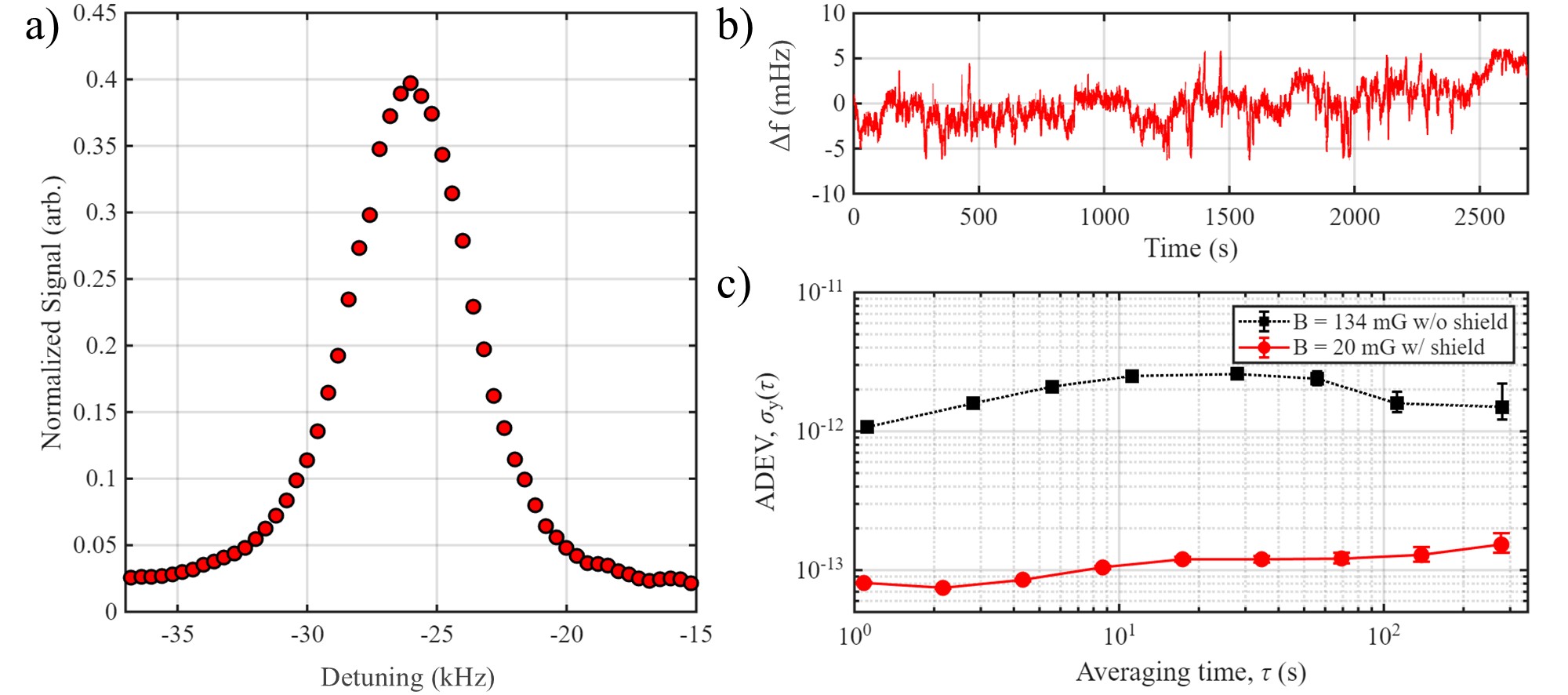}
    \caption{a) Microwave scan across the \(\left|F=1,\,m_\mathrm{F}=-1\right\rangle \rightarrow \left|F=2,\,m_\mathrm{F}=-1\right\rangle\) magnetically sensitive transition. b) Measured magnetic-field fluctuations mapped onto frequency fluctuations of the clock transition. c) Overlapping Allan deviation of the inferred clock-frequency fluctuations due to the second-order Zeeman shift at 134~mG without shielding (dotted black line with black squares) and 20~mG with shielding (solid red line with red circles).}
    \label{fig:2ndOrderZeemanADEV}
\end{figure}

We infer the quadratic Zeeman instability of the clock state by running a modified Rabi interrogation sequence, in which the optical-pumping light is changed from linear to circularly polarized and the repumper is turned off during optical pumping. This allows atoms to decay to the $F=1$ manifold and be preferentially pumped into the more magnetically sensitive $\lvert F=1,m_\mathrm{F}=-1\rangle$ state, which has a linear Zeeman sensitivity of $1.4~\mathrm{MHz/G}$. Rabi interrogation is used instead of Ramsey interrogation for this measurement, as a residual magnetic-field gradient across the atomic ensemble causes dephasing of the more magnetically sensitive state, washing out the Ramsey fringes. Further work is required to shim this gradient and enable Ramsey interrogation of the magnetically sensitive transition. A typical microwave scan of this transition is shown in Fig.~\ref{fig:2ndOrderZeemanADEV} a). The line centre of this transition is tracked, and the measured frequency deviations are converted to magnetic-field fluctuations using the known linear Zeeman sensitivity. The measured RMS magnetic-field fluctuation is $\approx 105~\mu\mathrm{G}$. The corresponding fractional frequency fluctuations of the clock transition are then inferred from

\begin{equation}
\frac{\delta f}{f_0}
=
\frac{2\beta_{m_\mathrm{F}=0}B\,\delta B}{f_0}.\label{fraqfreqBlimit}
\end{equation}

Here, $\beta_{m_\mathrm{F}=0}=575~\mathrm{Hz/G^2}$ is the quadratic Zeeman coefficient of the clock transition, $B$ is the applied bias magnetic field, $\delta B$ is the measured magnetic-field fluctuation, and $f_0$ is the unperturbed clock transition frequency. The resulting fractional frequency fluctuations of the clock transition are shown in Fig.~\ref{fig:2ndOrderZeemanADEV} b). An overlapping Allan deviation is then calculated from the fractional frequency time array, as shown in Fig.~\ref{fig:2ndOrderZeemanADEV} c). This corresponds to an inferred medium-term fractional frequency stability limit of approximately $1.5\times10^{-13}$.

The implementation of low-cost magnetic shielding therefore reduces the inferred contribution of second-order Zeeman fluctuations by more than an order of magnitude, removing one of the dominant medium-term stability limitations of the first-generation clock.

\section{Clock performance}
\begin{figure}[H]
    \centering
    \includegraphics[width=1\linewidth]{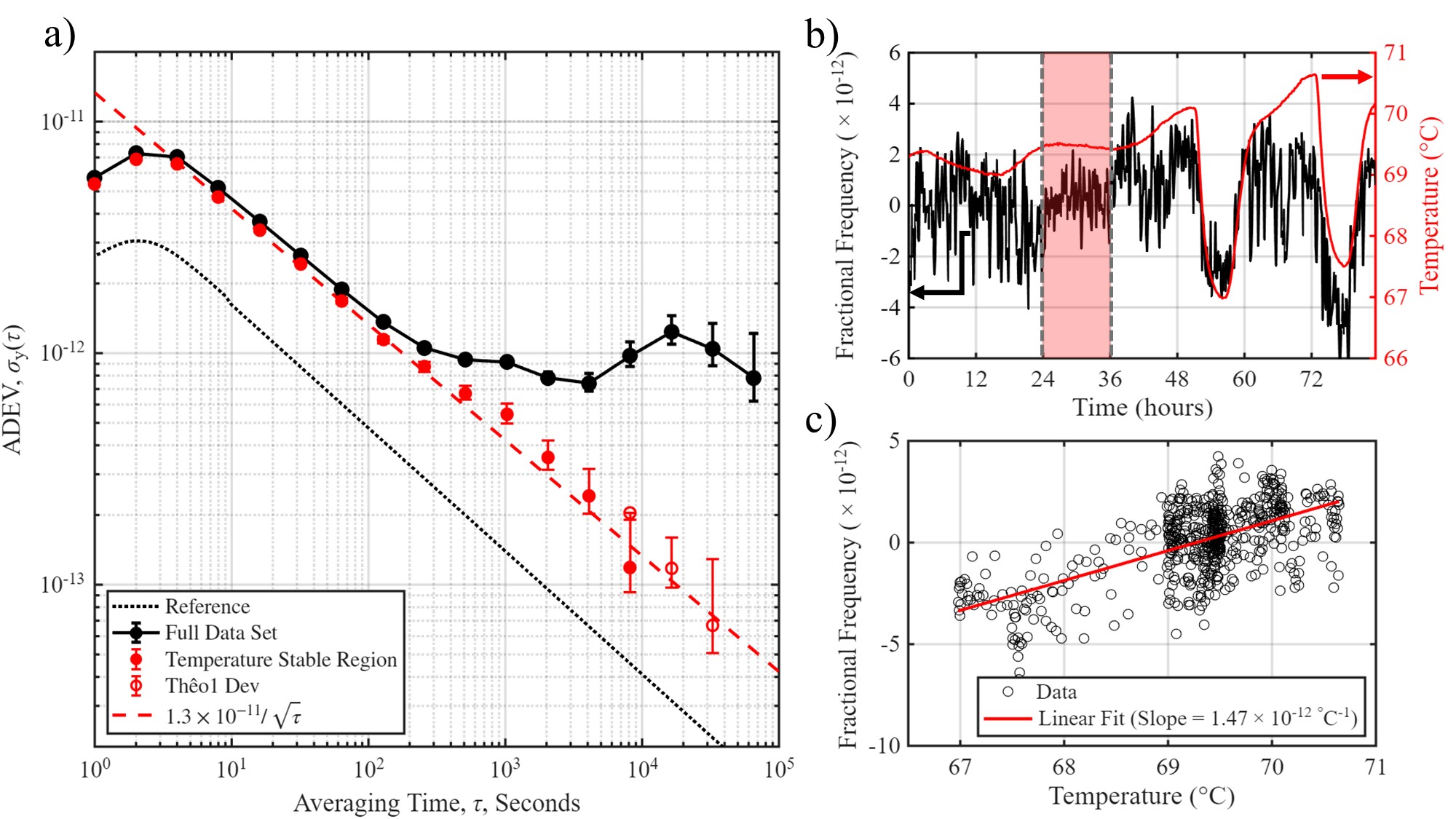}
    \caption{a) Overlapping Allan deviations. Black filled circles: Full measurement run over a 48 hour window. Red filled circles: most thermally stable section of the dataset. Red circles: Th\^{e}o1 Deviation of thermally stable section. Black dashed line: instability of reference.
    b) Fractional frequency averaged over 500~s, with the measured cavity temperature overlaid. Large induced temperature fluctuations are shown beyond 48 hours.
    c) Fractional frequency as a function of cavity temperature. The linear fit gives a temperature sensitivity of \(1.47\times10^{-12}~\mathrm{K}^{-1}\).
    }
    \label{fig:tempFig}
\end{figure}

To evaluate performance, the clock was locked continuously over a multiple day measurement. This duration was sufficient to assess the short-term stability and to identify the dominant medium-term limitation in the present system. The disciplined 10~MHz output from the clock was compared against a 5071B caesium frequency standard disciplined by UTC(NPL), using a Microchip 53100A phase noise analyser. The measured short-term stability was $\sigma_y(\tau)=1.3 \times 10^{-11}/\sqrt{\tau}$, as shown in Fig.~\ref{fig:tempFig} a). This is consistent with previous measurements performed under similarly low Rb vapour-pressure conditions~\cite{bregazzi2024lowvolume}. At averaging times greater that 100 seconds the measured stability departs from the expected $\tau^{-1/2}$ dependence, indicating an additional source of medium-term frequency instability. During the run, the cavity temperature was monitored using an in-vacuum thermocouple attached to the cavity. Figure~\ref{fig:tempFig} b) shows the fractional frequency averaged over 500~seconds, together with the measured cavity temperature. Temperature fluctuations were observed to coincide with corresponding changes in the locked clock frequency, indicating that the medium-term frequency instability is linked to variations in the cavity temperature.

To further investigate this effect, the cavity temperature was decreased by $3~\mathrm{K}$ while the locked clock frequency was monitored. A clear correlation between the cavity temperature and clock frequency was observed. The linear fit in Fig.~\ref{fig:tempFig} c) gives a fractional-frequency temperature sensitivity of $1.47\times10^{-12}~\mathrm{K}^{-1}$. This sensitivity is attributed to cavity pulling, which is a well-known temperature-dependent systematic in microwave atomic clocks and can degrade mid- to long-term clock stability~\cite{rossetto2011,bize2001,batori2023,almat2020}.

The cavity-pulling frequency shift is given by

\begin{equation}
\Delta\nu_{\mathrm{cp}}
=
-\frac{4}{\pi}
\frac{Q_c}{Q_a}
c(\theta)\Delta\nu_c,
\end{equation}

where $Q_c$ and $Q_a$ are the cavity and atomic quality factors, respectively, $c(\theta)$ is a function of the microwave pulse area, and $\Delta\nu_c$ is the detuning of the cavity resonance from the atomic transition. For this calculation, $c(\theta)=0.005$ is used based on the considerations given in Ref.~\cite{micalizio2010}. Using the measured cavity temperature coefficient of $\Delta\nu_c/\Delta T_{\mathrm{cav}}=-156\pm9~\mathrm{kHz/K}$ from Ref.~\cite{batori2023} gives an estimated fractional-frequency temperature sensitivity of approximately $4\times10^{-13}~\mathrm{K}^{-1}$, around a factor of four lower than the measured value. The source of this discrepancy is not yet understood and will be investigated in future work to identify effects not captured by the present model.

The uncontrolled temperature variations observed during clock operation point to changes in the local thermal environment of the microwave cavity. In the present apparatus, the cavity is mounted to the same in-vacuum structure that supports the MOT coils, which provide a local heat source, while the structure itself is mechanically connected to the outer wall of the vacuum chamber. Future iterations will improve the thermal mounting, isolation, and temperature control of the microwave cavity to mitigate this effect.

To estimate the stability achievable without large temperature variation, the most thermally stable section of the dataset was selected, indicated by the shaded region in Fig.~\ref{fig:tempFig} b). The overlapping Allan deviation for this section is also shown in Fig.~\ref{fig:tempFig} a), reaching a fractional frequency instability of \(1.2\times10^{-13}\) at \(\tau=10^{4}~\mathrm{s}\).

Due to the limited sample size, the Th\^{e}o1 deviation was used to provide improved statistical confidence at longer averaging times \cite{Theo1,fastlewis}, reaching a fractional frequency instability of \(6.7\times10^{-14}\) at \(\tau=3\times10^{4}~\mathrm{s}\).

\section{Outlook and Conclusion}\label{sec5}
We have demonstrated improved performance of a compact microwave atomic clock based on laser-cooled $^{87}\mathrm{Rb}$ atoms. This clock employs a gMOT for cooling and trapping atoms, and a compact additively manufactured LGR-type microwave cavity providing a scalable architecture by reducing both optical complexity and manufacturing requirements. Operating at high Rb density, a measured SNR of 370 at 1~Hz bandwidth corresponds to a short-term stability of $4.7 \times 10^{-12}/\sqrt{\tau}$, while the detection system itself is capable of achieving an SNR of up to 1100 for the same detected atom number. This discrepancy in SNR is attributed to frequency-to-amplitude noise conversion when using Ramsey-style interrogation. A low-cost, flexible magnetic shield has been implemented on the clock, reducing the Zeeman-limited medium-term instability due to magnetic field noise by an order of magnitude from $1.5 \times 10^{-12}$  to $1.5 \times 10^{-13}$ at $\tau =$100~s. In the mid-term, the measured clock performance indicates that cavity-temperature fluctuations are now the dominant limitation to medium-term stability, motivating improved thermal isolation and active temperature control in future designs. In future iterations of this clock, incorporating more flexible optical access to the cavity~\cite{galic2026} or implementing a MOT launch~\cite{lewis2022} could enable Ramsey interrogation times of up to 50~ms and 100~ms, respectively. The current system relies on loading atoms from background vapour, which typically results in relatively long cycle times and consequently increases the impact of the Dick effect~\cite{Santarelli} on short-term clock performance. Implementing a gMOT loaded from a 2D MOT has been demonstrated~\cite{cannon2026} and would reduce the cycle time, thereby mitigating the Dick effect and improving short-term stability.
Further redesign of the physics package is expected to reduce the overall package volume, with the microwave cavity potentially also serving as the ultra-high-vacuum chamber. Together, these developments provide a pathway towards a compact and scalable cold-atom microwave clock suitable for deployable precision timing applications.

\bmhead{Acknowledgments}
The authors would like to thank James P. McGilligan and Aidan S. Arnold for useful discussions and careful reading of the manuscript, and Rachel F. Offer for useful discussions.

\section*{Declarations}
\bmhead {Funding} This work was supported by EP/Z533178/1 :A U.K. Hub for Quantum Enabled Position, Navigation and Timing (QEPNT).
\bmhead {Conflicts of interest} The authors declare that they have no competing interests.
\bmhead {Data availability} The datasets supporting the findings of this study will be deposited in the University of Strathclyde PURE repository. A persistent DOI will be assigned and included prior to publication.
\bmhead {Author contributions} S.A.S performed experimental work, analysed data, co-wrote manuscript. S.D performed experimental work, analysed data and co-wrote manuscript. A.B performed experimental work and analysed data. R.C.D performed experimental work. O.S.B provided project guidance. G.M interpreted results and reviewed manuscript. C.A interpreted results and reviewed manuscript. E.R provided project guidance and acquired funding. P.F.G oversaw project, acquired funding, interpreted results, reviewed manuscript.

\newpage
\bibliography{sn-bibliography} 
\end{document}